# LE CHATELIER RESPONSE

B. Zilbergleyt
System Dynamics Research Foundation
livent@ameritech.net

The principle of Le Chatelier is one of the best-known fundamentals of chemical equilibrium. Being originally set up by its author qualitatively [1], it defines the direction where to the chemical system will move to accommodate a thermodynamic mismatch, resulted from changes of the system conditions: the stressed equilibrium system makes a dodge to another equilibrium state, where the stress is minimal. This is the simplest known example of self-organization in Nature. Le Chatelier principle is an extension of the Gauss' principle of least constraints [2,3] into chemical processes.

Since the time when the principle was initially worded, it is common to say that Le Chatelier 's principle is easier to illustrate than to state [4]. Well-known examples, used to explain how the principle works, are related to temperature or pressure changes. H. Le Chatelier [5] and R. Etienne [6] tried to move the principle towards the use of numbers, by offering simultaneously a moderation theorem for mole fractions in the form of an inequality. At about the same time, T. De Donder developed moderation theorems for thermodynamic affinity and basic thermodynamic parameters of state [7]. According to I. Prigogine and R. Defay [8], the moderation theorem for affinity $\delta A (d\xi/dt)_{p,T} > 0$ was the most general and correct thermodynamic formulation of Le Chatelier's principle. Indeed, it takes into account both key factors – thermodynamic affinity A as a general force and the reaction extent $\xi$, depending on the change.

Unfortunately, those efforts didn't turn Le Chatelier principle into a quantitative tool; likewise with Gauss' principle, it is still being employed exclusively for verbal explanations. Formulation of Le Chatelier's principle in a form of precise dependencies will be certainly of great help in chemical equilibrium analysis.

We will use the term "Le Chatelier's response" (or its abbreviation – LCR) to define the chemical system action elicited by an internal or external impact, the impact that creates constraints and forces the system to leave the existing equilibrium state for another one. Also, it makes sense to distinguish between disappearing, or "soluble" (soft) constraints, and persisting, or "insoluble" (harsh) constraints, that are put on the chemical system. The soluble constraints result in temporary stresses; they disappear traceless as soon as reaction attains new equilibrium state. That may correspond, e.g. to a one-time change in temperature, pressure or mole fractions. In isolated system, such a kind of changes in many cases may merely mean a system switch to a different section of the reacting space, defined by new chemical variables and relevant thermodynamic parameters.

The insoluble constraints do not disappear with the system adjustment to new equilibrium until the shifting force is ceased by an external hand. Insoluble constraints lead to minimal, but non-zero residual stresses at new equilibrium state. For instance, that kind of constraints occurs when the activity coefficients are changed; or in a system with two or more competing conjugated chemical reactions. It is noteworthy that the problem of soluble constraints is applicable to chemical reaction regardless the reacting system where it runs, even if nowhere (like in more than a century long tradition of classical chemical thermodynamics), while treatment of insoluble constraints strictly demands system approach.

The LCR mostly falls into a shift-force relationship. Our previous experience revealed a nearly universal shape of this dependence regardless the origin of thermodynamic force – usually it is relatively steep ascending curve with saturation (Fig.1, curve 1 on the left picture and dispersed points on the right picture; the right part is from [9]; all cases correspond to insoluble constraints).

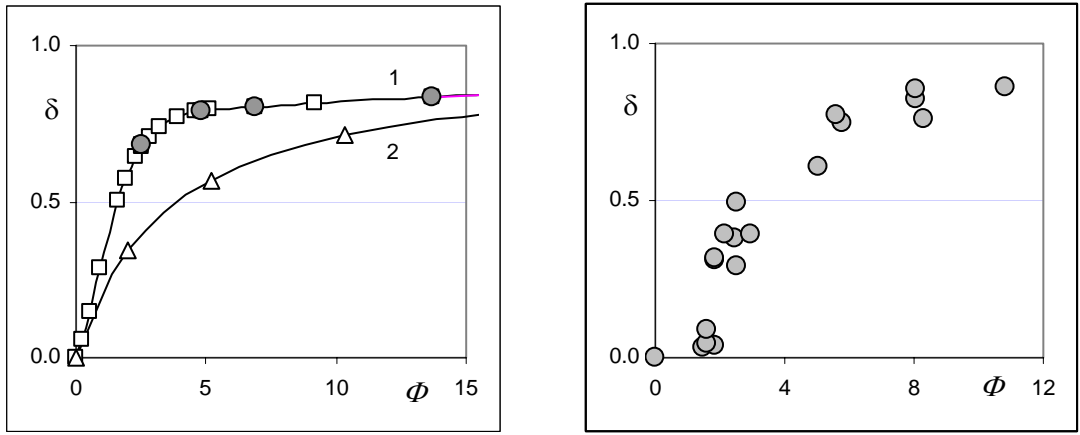

Fig.1. Shift-force graph. Left: 1–reaction (PbO•RO+S), chemical bound shift, RO is a non-reacting with S oxide: o–direct equilibrium simulation for double oxides, $\Phi = g/\Delta$, □ – imitational simulation with varying activity coefficients $\gamma$ of PbO, $\Phi = -\ln\gamma/\Delta$; 2 – pyrolysis of $CH_4$, shift is caused by temperature change. Right – shift is caused by chemical bounds, reaction (MeO•RO+S), various double-oxides, $\Phi = g/\Delta$.

To apprehend a soluble constraint, we have chosen the pyrolysis of methane in the temperature range of 273.15 to 1573.15 K at p=0.1MPa, resulting in its decay and release of hydrogen and soot. Using the HSC Chemistry, we found equilibrium compositions of the system. Taking a certain temperature as conditional base and by this means introducing the conditional base equilibrium state, we considered equilibrium at different temperatures as states, shifted from the base, and then have calculated the values of the shifting forces and appropriate shifts. The results are presented in Table 1 with the following notations: $\eta$ – thermodynamic equivalent of transformation of reaction $C+H_2=CH_4$ (in this case - the equilibrium amount of $CH_4$); $\Delta\xi$ (or just $\Delta$) – reaction extent (equals to unity at the base equilibrium); $\delta\xi$ (or just $\delta$) – reaction shift from thermodynamic equilibrium (equals to zero at the base). The negative value of $\Delta G/\Delta\xi$ is the thermodynamic affinity, in this case a driving thermodynamic force between two equilibrium states. In this example the base temperature was 973.15K. The system at any other temperature was considered perturbed with regard to the base and subject to LCR commensurate with the extent of deviation from the base. Two values in Table 1 were calculated by formulas (see [10]): $\Delta = \eta/0.193$, where 0.193 is the base $\eta$ value, $\delta = 1-\Delta$. In our definition, $\Delta$ indicates system proximity to thermodynamic equilibrium, in this case to the base, taking its highest value of unity at that point. This data is turned into a graph in Fig.1, left, curve 2. To move the starting point of the curve to zero, the reference frame was shifted towards negative values of $\Delta G$ by 16.280 kJ/mol. All LCR graphs in Fig.1 apparently are similar.

Table 1.
Parameters of Chemical Equilibrium for Pyrolysis of $CH_4$ (initial amount 1 mol).

| T, K | $\eta$ | $\Delta\xi$ | $\delta\xi$ | $\Delta G$, kJ/mol | $\Delta G/\Delta\xi$, kJ/mol |
|---|---|---|---|---|---|
| **973.15** | 0.193 | 1.000 | 0.000 | 16.280 | 16.280 |
| 1023.15 | 0.127 | 0.658 | 0.342 | 21.711 | 32.994 |
| 1073.15 | 0.084 | 0.433 | 0.567 | 27.164 | 62.786 |
| 1123.15 | 0.056 | 0.289 | 0.711 | 32.635 | 113.080 |
| 1173.15 | 0.038 | 0.196 | 0.804 | 38.121 | 194.125 |
| 1223.15 | 0.026 | 0.137 | 0.863 | 43.617 | 318.867 |
| 1273.15 | 0.019 | 0.097 | 0.903 | 49.121 | 504.274 |
| 1323.15 | 0.010 | 0.053 | 0.947 | 54.631 | 1033.704 |
| 1373.15 | 0.007 | 0.038 | 0.962 | 60.144 | 1592.290 |
| 1423.15 | 0.005 | 0.028 | 0.972 | 65.659 | 2355.425 |
| 1473.15 | 0.004 | 0.021 | 0.979 | 71.175 | 3391.796 |
| 1523.15 | 0.003 | 0.016 | 0.984 | 76.690 | 4743.965 |
| 1573.15 | 0.002 | 0.013 | 0.987 | 82.203 | 6475.583 |

An attempt to find analytical relations between the shift and the shifting force was done in [10] with regards to an open chemical system under insoluble constraints and with only one chemical reaction allowed to run. To account for possible complexity in the shift-force formula, we define for LCR the term $\rho_j$ as a function of the reaction shift from thermodynamic equilibrium, $\delta_j$

(1) $$\rho_j = f(\delta_j),$$

and corresponding LCR-force equation as

(2) $$\rho_j = -(1/\alpha) F_j,$$

where the perturbing force $F_j$ has the dimension of energy gradient at a dimensionless reaction extent; $\rho_j$ is dimensionless; $\alpha$ has the same dimension as the force. Following the derivation procedure from [10], we will receive at constant p and T a new equation for the reduced by RT change of Gibbs' free energy of the system at chemical equilibrium under constraint as

(3) $$\ln[\Pi_j(\eta_j,0)/\Pi_j(\eta_j,\delta_j)] - \tau_j\varphi(\delta_j) = 0,$$

where $\delta_j = 1-\Delta_j$ and $\Delta_j$ is reaction extent, $\Pi_j(\eta_j,\delta_j)$ is a regular mole fractions product in chemical equilibrium. Quite obviously at $\delta_j=0$ we have the equilibrium constant $K_{eq}=\Pi_j(\eta_j,0)$, and $\tau_j\varphi(\delta_j)$ is a non-classical term, defining the constraint. The factor $\tau_j$ is the *chaotic temperature*, the ratio of the product of coefficient $\alpha$ by corresponding Onsager coefficient (see [10] for details) and RT. The $\tau_j$ is a parameter, identical to the growth factor in the theory of bio-populations [11]; behavior of the open chemical system strongly depends on that value. Chemical equilibrium, described by equation (3) corresponds to a certain shifting force; it can be found dividing any term of this equation by the reaction extent $\Delta_j$.

To find $\varphi(\delta_j)$, we write the response down as a sum of various powers of the reaction shift

(4) $$\rho_j = \Sigma \nu_j \delta_j^n,$$

where $\delta_j \leq 1$, and the series power n may or may not start with zero and may or may not tend to infinity. The powers of $\delta_j$ in formula (4) may have different, but so far unknown weights $\nu_j$; we will equate them to unity. With this remark, contents of the n-series define the value of the second term as follows:

- if $n = \{0, ..., \to \infty\}$ we have $\rho_j = 1/\Delta_j$, and $\varphi(\delta_j) = 1$;
- if $n = \{1, ..., \to \infty\}$ we have $\rho_j = \delta_j/\Delta_j$, and $\varphi(\delta_j) = \delta_j$;
- if the $\delta_j^n$ – sum starts with n=1 and its upper limit is restricted to relatively small n values, we have $\rho_j = \delta_j + \delta_j^2 + ... + \delta_j^n$, and $\varphi(\delta_j) = \delta_j(1-\delta_j^n)$;
- in a particular case, if the sum contains only term with n=1, $\rho_j = \delta_j$, and we have $\varphi(\delta_j) = \delta_j(1-\delta_j)$.

When the stressed system shifts to new equilibrium state, in case of soluble constraints the second term of equation (3) vanishes due to $\delta_j=0$, otherwise taking the minimal possible value. The case of $\varphi(\delta_j) = 1$ doesn't satisfy any of the equations, and should be eliminated from consideration. The last of the above listed cases leads to the logistic equation for Gibbs' free energy change in an open system as

(5) $$\ln[\Pi_j(\eta_j,0)/\Pi_j(\eta_j,\delta_j)] - \tau_j\delta_j(1-\delta_j) = 0.$$

Graphical solutions to equation (5), known as bifurcation diagrams, depict evolution of chemical equilibrium, driven sequentially through the areas of true and open equilibrium up to bifurcations and chaos when $\tau_j$ increases [10].

We do not know any reported occurrence of bifurcations (or anything worse) when the perturbed chemical system under soluble constraints gains new equilibrium state. So, the best option for the soluble constraints is

(6) $\qquad\qquad\qquad\qquad\qquad\qquad \ln[\Pi_j(\eta_j,0)/\Pi_j(\eta_j,\delta_j)] - \tau_j\delta_j = 0.$

Taking into account that one cannot find the $\tau_j$ value from the experiment (including direct computer simulation experiment), we have compared appropriate curves for both shapes of the second term, obtained by computer experiment, and data for a soluble constraint on the $C+H_2=CH_4$ reaction from Table 1 (Fig.2). We would like to focus the reader's attention on 2 features of the curves in Fig.2. First, solution curves to both equations (5) and (6) very well match each other prior to bifurcation point. Second, the curve 3 for a soluble constraint actually follows the same pattern, and its fit in two previous curves is just a matter of scaling. A direct proof of absence of bifurcations in case of equation (6) gives curve 4 in Fig.3, obtained by iteration technique. Curves 1, 2, 3 in Fig.3 originated from equation (5) with $\rho_j = \Sigma \delta_j^n$ and $\varphi(\delta_j)=\delta_j(1-\delta_j^n)$, where initially n=1 and the limit values of n were 1, 2, and 3. Interesting enough that due to $\delta_j<1$, the contribution of the $\delta_j^n$ member declines as n increases, and eventually $(1-\delta_j^n)$ tends to unity. The curves keep the same shape for $\tau_j\delta_j(1-\delta_j^n)$ with well expressed degradation in the bifurcation diagrams toward the simple ascending curve 4 of Fig.3.

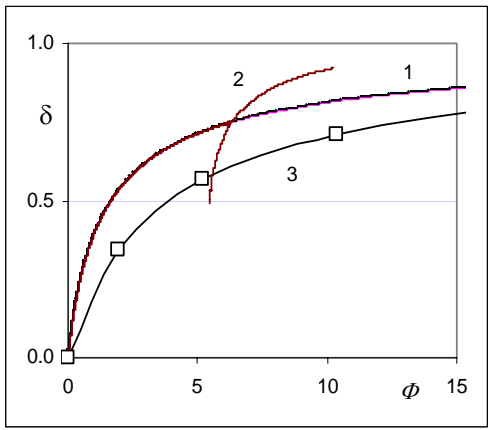
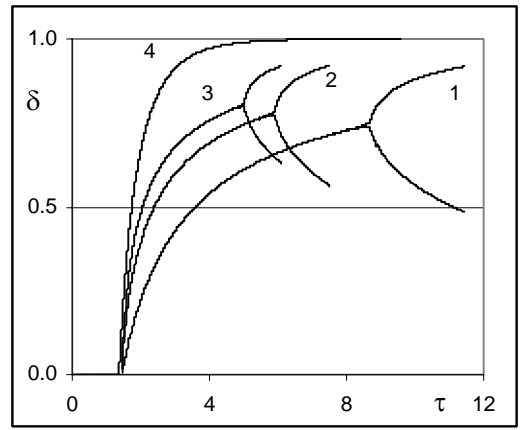

Fig.2. Shift-force relations for logistic equations with the 2d term: 1- $\tau_j\delta_j$, 2-$\tau_j\delta_j(1-\delta_j)$, and for reaction $C+H2=CH4$ (curve 3).

Fig.3. Solution diagrams for equations (6) (curves 1, 2, 3) and (5) (curve 4).

The delay area on the abscissa, corresponding to the area of thermodynamic equilibrium converges into point when the curves in Fig.3 are transformed into shift-force coordinates.

Our observation prompts us to assume that equation (6) may well describe what happens to the chemical system under a soluble constraint. It looks as if the more complicated expression for $\rho_j$ gives simpler results and vise versa. Consideration of possible internal degrees of freedom of the chemical system may provide some kind of an explanation for this result. Indeed, bifurcations occur at the point where the thermodynamic branch gets unstable; in a common language, this is a point where the external impact against the system, which forces it to decrease the reaction extent, becomes an intolerable burden. Higher powers of $\delta_j^n$ in LCR indicate more complicated internal organization with more internal degrees of freedom, and more flexible resilience to external impact; such a system may be less prone to splitting its states. Then, at $n\to\infty$ we get a limit like curve 4 in Fig.4 corresponding to rather "soft" impact like a change of temperature, while a stronger external impact may suppress the degrees of freedom at higher powers of n, bringing $\varphi(\delta_j)$ down to $\delta_j(1-\delta_j^n)$ with limited n value and even to $\delta_j(1-\delta_j)$ – as in the case we investigated in [10] and in some other previous publications.

To conclude, one should mention that in many applications logistic equations are used to create a rough picture of a system behavior with bifurcations and to allow for verbal explanations. In our previous works (see [10]) we set as an objective to create the system domain of states using logistic equations to analyze chemical systems with more possible precision. Numerical presentation of Le Chatelier's principle was a key idea to success in the thermodynamics of chemical systems, unifying the equilibrium and non-equilibrium branches on a unique basis in the treatment of chemical equilibrium.

As always, the author is grateful to his very convincing editor, Dr. Nolte from the EditAvenue.com.